\documentclass[preprint,sort&compress, 5p,times]{elsarticle}

\usepackage{lipsum}
\makeatletter
\def\ps@pprintTitle{%
 \let\@oddhead\@empty
 \let\@evenhead\@empty
 \def\@oddfoot{}%
 \let\@evenfoot\@oddfoot}
\makeatother

\usepackage{float,ulem}
\usepackage{graphicx,epsfig,epstopdf,amsmath,amssymb}
\usepackage{xcolor}
\usepackage[colorlinks=true,linkcolor=blue,citecolor=blue,urlcolor=blue]{hyperref}
\usepackage{caption}
\usepackage{subcaption}

\newcommand{\be}{\begin{equation}}
\newcommand{\ee}{\end{equation}}
\newcommand{\ba}{\begin{eqnarray}}
\newcommand{\ea}{\end{eqnarray}}
\newcommand{\nn}{\nonumber\\}

\begin{document}

\title{Causality and stability in relativistic hydrodynamic theory - a choice to be endured}

\author[1,2]{Sayantani Bhattacharyya}
\ead{sayanta@niser.ac.in}

\author[2]{Sukanya Mitra}
\ead{sukanya.mitra@niser.ac.in}

\author[2]{Shuvayu Roy}
\ead{shuvayu.roy@niser.ac.in}
\address[1]{School of Mathematics,
University of Edinburgh, United Kingdom.}

\address[2]{School of Physical Sciences, National Institute of Science Education and Research, An OCC of Homi Bhabha National Institute, Jatni-752050, India.}

\begin{abstract}
In this work, it has been indicated that the key features requisite for preserving causality and stability of the popularly existing relativistic hydrodynamic theories, can be translated into each other. It has been shown here, that a generic `fluid frame transformation' including all orders of gradient corrections can recast a stable-causal hydrodynamic theory that (i) only includes fundamental fluid variables (velocity and temperature) but requires to be in a general hydrodynamic frame other than the Landau or Eckart, to a theory that is (ii) pathology free in Landau frame but needs newer degrees of freedom. Since frame choice provides the first principle field definitions and degrees of freedom indicate the number of conserved quantities, the causality and stability of a theory seem to require a consensus between the two unless the derivative correction goes to infinity. The key finding of this work is to indicate a connection between these different formalisms, that lead to a causal and stable hydrodynamic theory.
\end{abstract}
\maketitle

\section{Introduction}
What requires a hydrodynamic theory to be physically acceptable? The widely accepted
benchmark criteria to pass it as a reliable theory are that the governing equations must predict, (a) causal wave propagation, i,e,
the signals do not propagate faster
than the light and, (b) stability against small perturbations around the global thermodynamic equilibrium such that they do not grow indefinitely over time. Apart from these two imperative criteria, a hydrodynamic theory prefers certain desirable features. First, (i) they are defined only in terms of the dynamical variables, such as temperature ($T$), fluid velocity ($u^{\mu}$), and chemical potential ($\mu$), that are related to some conserved quantities characterising the equilibrium. Second, (ii) even out of equilibrium, the fluid fields are well defined in terms of microscopic operator. The last one is that (iii)  the final differential equations for the fluid variables must contain a finite number of derivatives. (i) and (ii) are needed to build an unambiguous contact between the fluid variables and the quantities that are experimentally measurable. The third feature (iii) is a must to have computationally tractable evolution equations.

In reality, the journey of relativistic dissipative hydrodynamics has not been much smooth and this desirable holy trinity is far from being realized. It turns out if we choose the standard Landau or Eckart frame to define the fluid variables precisely and in a physically measurable way, in order to maintain the causality and the stability we have to have either infinite number of derivatives or new `non-fluid' degrees of freedom. The first issue appears when the relativistic first-order theory at both Landau and Eckart frames \cite{Landau,Eckart:1940te}, famously known as the Navier-Stokes (N-S) theory, ends up with superluminal signal propagation and thermodynamic instability. Next, the higher (but finite) order derivative theories like the Muller-Israel-Stewart (MIS) theory
\cite{Muller,Israel:1976efz,Israel:1976tn,Israel:1979wp} are introduced. Here apart from the basic fluid variables ($T,\mu,u^{\mu}$) the dissipative fluxes that are
not associated with any thermodynamic conserved charges (and with no equilibrium counterparts) are required to be promoted as new variables, with additional equations of motion to ensure the stability-causality at Landau or Eckart frame \cite{Hiscock:1983zz,Hiscock:1985zz,Olson:1990rzl,Denicol:2012cn,Baier:2007ix,Muronga:2001zk}. Recently in \cite{Bemfica:2017wps,Bemfica:2019knx,Bemfica:2020zjp,Kovtun:2019hdm,Hoult:2020eho,Hoult:2021gnb,Rocha:2022ind,Biswas:2022cla,Biswas:2022hiv} another interesting way of constructing a causal-stable fluid theory (popularly known as the Bemfica-Disconzi-Noronha-Kovtun (BDNK) theory) has been introduced which has a finite number of derivatives but no extra `non-fluid' type variables. But the definition of the fluid variables away from equilibrium have not been fixed in this formalism though it is easy to see the theory is  pathology-free only when it is not in  the Landau (or Eckart) frame.

In this work, the general frame BDNK stress tensor has been rewritten in the Landau frame by redefining the fluid variables ($T$ and $u^{\mu}$) up to the linear order of derivative corrections. The redefinition takes the finite ordered BDNK formalism to a system of equations that either has an infinite number of derivatives \cite{Mitra:2023ipl} or has to introduce new `non-fluid' variables as in the MIS theory. Our work indicates that there is a tension between the three desirable features of hydrodynamics (i), (ii) and (iii). To maintain causality and stability we have to give up at least one of them. BDNK formalism gives up (ii), MIS gives up (i). Our frame-transformed BDNK theory either gives up (iii) by having terms with infinite number derivatives or gives up (i) by introducing new non-fluid variables and equations just like MIS, but we maintain (ii).

 However, the method of `integrating in' new `non-fluid' variables is not unique and one can adopt different approaches, two of which have been discussed here.
 The primary objective of this letter is to suggest a connection between these two major formalisms (MIS and BDNK) which so far have served as the most established relativistic dissipative hydrodynamic theories. Previously, in \cite{Das:2020fnr,Das:2020gtq} a very similar dynamics (temperature evolution with time) predicted by MIS and BDNK theory has been noted for boost-invariant, Bjorken expanding systems. Here, to the best of our knowledge, for the first time, the connection has been
attributed to a physical origin with the required mathematical framework. In this context, we mention the analysis in \cite{Dore:2021xqq} where the  MIS and the BDNK type formalism have been connected with field redefinition more from a microscopic point of view. Whereas, in our analysis, we are completely agnostic about the microscopic descriptions or statistical interpretation of these field redefinitions. As a result, we could find more than one way (in fact, in principle, there should be just an infinite number of ways) of `integrating in' the non-fluid variables for the same BDNK theory but recast in Landau frame.

We adopt the metric signature $g^{\mu\nu}=(-1,1,1,1)$ and we work in natural units $c=k_B=1$.
%-----------------------------------------------
%-----------------------------------------------
\section{`Fluid frame' transformation of BDNK theory}
With conformal symmetry and no conserved charges, the energy-momentum tensor ($T^{\mu\nu}$) in the BDNK theory is defined as \cite{Bemfica:2017wps,Bemfica:2019knx,Bemfica:2020zjp,Kovtun:2019hdm,Hoult:2020eho,Hoult:2021gnb},
\begin{align}
&T^{\mu\nu}=(\varepsilon +{\cal{A}})\left[u^{\mu}u^{\nu}+\frac{\Delta^{\mu\nu}}{3}\right]+\left[u^{\mu}Q^{\nu}+u^{\nu}Q^{\mu}\right]-2\eta\sigma^{\mu\nu},
\nn
&{\cal{A}}={\chi}\left[3\frac{DT}{T}+\nabla_{\mu}u^{\mu}\right],~~
Q^{\mu}=\theta\left[\frac{\nabla^{\mu}T}{T}+Du^{\mu}\right].
 \label{BDNK}
 \end{align}
Here, $\eta$ is the shear viscous coefficient and
$\chi,\theta$ are the first-order field correction coefficients of BDNK theory. The used notations read,
$D= u^{\mu}\partial_{\mu}$, $\nabla^{\mu}=\Delta^{\mu\nu}\partial_{\nu}$,
$\sigma^{\mu\nu}=\Delta^{\mu\nu}_{\alpha\beta} \partial^{\alpha}u^{\beta}$ with $\Delta^{\mu\nu\alpha\beta}=\frac{1}{2}\Delta^{\mu\alpha}\Delta^{\nu\beta}+\frac{1}{2}\Delta^{\mu\beta}\Delta^{\nu\alpha}
-\frac{1}{3}\Delta^{\mu\nu}\Delta^{\alpha\beta}$ and
$\Delta^{\mu\nu}=g^{\mu\nu}+u^{\mu}u^{\nu}$.
For a conformal system the energy density $\varepsilon$, pressure $P$ and $T$ are connected as, $\varepsilon\sim T^4$, $(\varepsilon+P)=4\varepsilon/3$ and $D\varepsilon/(\varepsilon+P)=3DT/T$.
Eq.\eqref{BDNK} produces stable-causal modes only with non-zero values of $\theta$ and $\chi$. The neatness of this formalism lies in not requiring any additional degrees of freedom other than temperature and velocity and Eq.\eqref{BDNK} also shows that the theory is local in fluid variables both spatially and temporally. However, as mentioned before, unlike the MIS theory, the definitions of the fluid velocity and the temperature are not fixed here in terms of stress tensor or any other microscopic operator. Here, we would like to redefine the velocity and the temperature in a way so that the stress tensor, expressed in terms of these redefined fluid variables, satisfies the Landau frame condition. We define the BDNK variables $u^\mu$ and $T$ in Eq.\eqref{BDNK} to relate with the Landau frame velocity $\hat u^\mu$ and temperature $\hat T$ in the following fashion,
\be
u^\mu = \hat u^\mu +\delta u^\mu,~~~~T = \hat T+ \delta T~.
\label{shift}
\ee
The shift variables $\delta u^\mu$ and $\delta T$, which are non-trivial functions of $\hat u^\mu$ and $\hat T$, are small enough to be treated only linearly but encompass all orders of gradient corrections. Once we impose the Landau gauge condition $T^{\mu}_\nu \hat u^\nu=-\hat\varepsilon \hat u^\mu$ after substituting \eqref{shift} in the BDNK stress tensor \eqref{BDNK}, it reduces to the following set of coupled linear PDEs for the shift variables,
\begin{align}
&\frac{\delta T}{\hat{T}}+\tilde{\chi}\left[\frac{\hat{D}\hat{T}}{\hat{T}}
+ \frac{\hat{\nabla}_{\mu}\hat{u}^{\mu}}{3}\right]+\tilde{\chi}\left[\frac{\hat{D}\delta{T}}{\hat{T}}+\frac{\hat{\nabla}_{\mu}\delta{u}^{\mu}}{3}\right]=0~,
\nn
 &\delta u^{\mu}+\tilde{\theta}\left[\hat{D}\hat{u}^{\mu}+\frac{\hat{\nabla}^{\mu}\hat{T}}{\hat{T}}\right]+\tilde{\theta}\left[\hat{D}\delta{u}^{\mu}+\frac{\hat{\nabla}^{\mu}\delta{T}}{\hat{T}}\right]=0~,
 \label{shifteqn}
\end{align}
where $\hat{D}= \hat{u}^{\mu}\partial_{\mu}$, $\hat{\nabla}^{\mu}=\hat{\Delta}^{\mu\nu}\partial_{\nu}$ with $\hat{\Delta}^{\mu\nu}=g^{\mu\nu}+\hat{u}^{\mu}\hat{u}^{\nu}$ and $\tilde{x}=x/(\varepsilon+P)$.
In deriving equations \eqref{shifteqn}, the nonlinear terms in $\delta u^{\mu}$ and $\delta T$ are ignored. This linearization simplifies the analysis so that we can analytically have an `all-order' (in derivatives)
formula for both the field corrections (the solutions of $\delta T$ and $\delta u^{\mu}$ from \eqref{shifteqn}) in the new frame.

The fact that Eq.\eqref{shifteqn} are not algebraic but a set of  a differential equations for the shift variables $\delta T$ and $\delta u^\mu$ already hints at the emergence of `MIS type nonlocality'  even in the BDNK theory. At the linearized level, we have done the field redefinition in two different representations. In one case, we summed only the time derivatives up to the infinite order, leading to a set of equations that look nonlocal in time (with the time derivative appearing in the denominator) but local in space. In the second case, we summed both the time and the space derivatives, leading to a full nonlocal redefinition of the fluid variables. In either case, these nonlocalities (derivatives appearing in the denominator) could be absorbed by introducing new `non-fluid' variables, identical to defining new degrees of freedom in the MIS theory. These two different methods of introducing `non-fluid' degrees of freedom to make BDNK a local theory in the Landau frame are respectively described in the following.
%-----------------------------------------------
%===============================================
\section{Frame transformation order by order}
We assume that $\delta u^\mu$ and $\delta T$ admit the following infinite series expansion,
\be
 \delta u^\mu =\sum_{n=1}^{\infty}\delta u_n^{\mu},~~~~~~\delta T =\sum_{n=1}^{\infty}\delta T_n~,
 \label{shift-order}
\ee
where the subscript $n$ denotes the order correction in terms of derivative expansion. Substituting \eqref{shift-order} in the PDEs \eqref{shifteqn}, the solutions can be found in terms of the following recursive relations,
\begin{align}
&\delta T_1=-\tilde{\chi}\left[\frac{\hat{D} \hat{T}}{\hat{T}}+\frac{1}{3}\hat{\nabla}_{\mu}\hat{u}^{\mu}\right],~~~\delta u_1^{\mu}=-\tilde{\theta} \left[\frac{\hat{\nabla}^{\mu}\hat{T}}{\hat{T}}+\hat{D}\hat{u}^{\mu}\right],\nn
&\delta T_n=-\tilde{\chi}\left[\frac{1}{\hat{T}}\hat{D}\delta T_{n-1}+\frac{1}{3}\hat{\nabla}_{\mu}\delta u^{\mu}_{n-1}\right]~~~~\text{for}~n\geq 2~,
\nn
&\delta u_n^{\mu}=-\tilde{\theta} \left[\frac{1}{\hat{T}}\hat{\nabla}^{\mu}\delta T_{n-1}+\hat{D}\delta u^{\mu}_{n-1}\right]~~~~\text{for}~n\geq 2~.
\label{shift-soln}
\end{align}
Eq.\eqref{shift-soln} provides the successive field corrections up to any desired order. Rewriting the energy-momentum tensor \eqref{BDNK} in terms of the new fluid variables $\hat{u}^{\mu}$, $\hat{\varepsilon} (\hat{T})$ and substituting the recursive solution for $\delta u^\mu_n$ and $\delta T_n$ as given in \eqref{shift-soln}, the energy density correction $({\cal{A}})$ and energy-flux or momentum flow $Q^{\mu}$ vanish in $T^{\mu\nu}$ as expected in the Landau frame, and one finally has the following energy-momentum tensor upto all order,
\begin{align}
T^{\mu\nu}=&\hat{\varepsilon}\left[\hat{u}^{\mu}\hat{u}^{\mu}+\frac{1}{3}\hat{\Delta}^{\mu\nu}\right]
-2\eta\left[\hat{\sigma}^{\mu\nu}+\sum_{n=1}^{\infty}\partial^{\langle\mu}\delta u_n^{\nu\rangle}\right],
\label{BDNK1}
\end{align}
with $\hat{\sigma}^{\mu\nu}=\partial^{\langle\mu}\hat{u}^{\nu\rangle}=\hat{\Delta}^{\mu\nu}_{\alpha\beta}\partial^{\alpha}\hat{u}^{\beta}$. In order to construct $T^{\mu\nu}$ in this new 'fluid frame' given by Eq. \eqref{BDNK1}, we need to evaluate $\delta u_n^{\mu}$ from \eqref{shift-soln}. Once the explicit expressions of $\delta u^\mu_n$ and $\delta T_n$ are estimated order by order, a very nice pattern emerges, which can be used to sum the infinite series given in \eqref{shift-order} to get an all-order expression of the shift variables.
In the expressions of $\delta u_n^{\mu}$ and $\delta T_n$, with increasing order of $n$, terms with higher order of the spatial gradients on $T$ and $u^{\mu}$ are appearing systematically and also the order of the temporal gradient on each such spatial gradient term chronologically increases. This increase of temporal derivatives is observed to follow a particular pattern such that they can be clubbed together into products of infinite sums. The infinite sums over the time derivative can be encompassed in a closed form following
the relaxation operator-like terms to appear in the denominator of the shift fields, giving rise to pole-like structures. The explicit expressions of $\delta T_n$ and $\delta u^{\mu}_n$ and the details of the infinite summation techniques can be found in a separate, parallel manuscript.

Putting this resummed velocity correction ($\delta u^{\mu}_n$ all orders) in Eq.\eqref{BDNK1} we have the all order frame transformed BDNK stress tensor in Landau frame as,
\begin{align}
T^{\mu\nu}=\hat{\varepsilon}\left[\hat{u}^{\mu}\hat{u}^{\nu}+\frac{1}{3}\hat{\Delta}^{\mu\nu}\right]+\hat{\pi}^{\mu\nu}~,
\label{BDNK00}
\end{align}
with $\hat{\pi}^{\mu\nu}=-2\eta\left[\hat{\sigma}^{\mu\nu}+\sum_{n=1}^{\infty}\partial^{\langle\mu}\delta u_n^{\nu\rangle}\right]$, now summed over temporal derivatives under the all order frame transformation as the following,
\begin{align}
\hat{\pi}^{\mu\nu}=&-2\eta\bigg[
\frac{\hat{\nabla}^{\langle\mu}\hat{u}^{\nu\rangle}}{(1+\tilde{\theta}\hat{D})}
+\frac{(-\tilde{\theta})}{(1+\tilde{\theta}\hat{D})}\frac{\frac{1}{\hat{T}}\hat{\nabla}^{\langle\mu}\hat{\nabla}^{\nu\rangle}\hat{T}}{(1+\tilde{\chi}\hat{D})}\nn
+&\frac{(-\tilde{\theta})}{(1+\tilde{\theta}\hat{D})^2}\frac{(-\frac{1}{3}\tilde{\chi})}{\left(1+\tilde{\chi}\hat{D}\right)}\hat{\nabla}^{\langle\mu}\hat{\nabla}^{\nu\rangle}\hat{\nabla}\cdot\hat{u}\nn
+&\frac{(-\tilde{\theta})^2}{(1+\tilde{\theta}\hat{D})^2}\frac{\left(-\frac{1}{3}\tilde{\chi}\right)}{(1+\tilde{\chi}\hat{D})^2}
\frac{1}{\hat{T}}\hat{\nabla}^{\langle\mu}\hat{\nabla}^{\nu\rangle}\hat{\nabla}^2\hat{T}\nn
+&\frac{(-\tilde{\theta})^2}{(1+\tilde{\theta}\hat{D})^3}\frac{\left(-\frac{1}{3}\tilde{\chi}\right)^2}{(1+\tilde{\chi}\hat{D})^2}
\hat{\nabla}^{\langle\mu}\hat{\nabla}^{\nu\rangle}\hat{\nabla}^2\hat{\nabla}\cdot\hat{u}+\cdots
\bigg].
\label{BDNK11}
\end{align}
Note that for each increasing spatial gradient, the temporal gradient resulting from the infinite sum also increases in the denominator, such that they exactly balance each other. This condition has been mentioned in \cite{Hoult:2023clg}  as a necessary condition of causality. So this infinite derivative summation of the all-order frame transformation pursued so far has not
tampered with the causality of the original BDNK theory given in \eqref{BDNK}.

We see that Eq. \eqref{BDNK11} is just a formal solution as it has derivatives (in time) in the denominator. Such an expression really makes sense in the space of frequencies rather than in real time. However, what this indicates is a nonlocality in time (or integration over time). Such nonlocalities can be recast into a local set of equations by introducing new 'non-fluid' variables just like the MIS theory \cite{Romatschke:2009im}. Like $\pi^{\mu\nu}$ (shear viscous flow) in MIS theory, these 'non-fluid' degrees of freedom (vanish at any state of global equilibrium and therefore, are not extensions of any conserved charges) should approach a vanishing value in a `relaxation type' equation. Here, the relaxation time scales are provided by the poles of the infinite sum of temporal derivatives appearing in the denominator of Eq.\eqref{BDNK11}. However, unlike the MIS theory, here, after completing the infinite sum in the temporal derivatives, the degree of the pole increases ad infinitum along with more and more spatial derivatives in the numerator. This indicates an infinite number of non-fluid degrees of freedom in a nested series of `relaxation type' equations. The following set of infinitely many equations provides a local theory both in space and time at the Landau frame which is equivalent to the BDNK theory, at least with respect to linearized perturbations around equilibrium :
\begin{align}
 &\partial_\mu T^{\mu\nu}=0~,~~~~~~~~
 T^{\mu\nu}=\hat{\varepsilon}\left[\hat{u}^{\mu}\hat{u}^{\nu}+\frac{1}{3}\hat{\Delta}^{\mu\nu}\right]+\hat{\pi}^{\mu\nu},\nn
 &(1+\tilde\theta\hat D)\hat{\pi}^{\mu\nu}=-2\eta\hat{\sigma}^{\mu\nu}+\rho_1^{\mu\nu},\nn
 &(1+\tilde\chi\hat D)\rho^{\mu\nu}_1=(-2\eta)(-\tilde{\theta})\frac{1}{\hat{T}}\hat{\nabla}^{\langle\mu}\hat{\nabla}^{\rangle\nu} \hat{T}+\rho_2^{\mu\nu},\nn
 &(1+\tilde\theta\hat D)\rho^{\mu\nu}_2=(-2\eta)(-\tilde{\theta})\left(-\frac{\tilde{\chi}}{3}\right)\hat{\nabla}^{\langle\mu}\hat{\nabla}^{\rangle\nu} \hat{\nabla}\cdot\hat{u}+\rho_3^{\mu\nu},\nn
 &(1+\tilde\chi\hat D)\rho^{\mu\nu}_3=(-2\eta)(-\tilde{\theta})^2\left(-\frac{\tilde{\chi}}{3}\right)\frac{1}{\hat{T}}\hat{\nabla}^{\langle\mu}\hat{\nabla}^{\rangle\nu} \hat{\nabla}^2 \hat{T}+\rho_4^{\mu\nu},\nn
 &(1+\tilde\theta\hat D)\rho^{\mu\nu}_4=(-2\eta)(-\tilde{\theta})^2\left(-\frac{\tilde{\chi}}{3}\right)^2\hat{\nabla}^{\langle\mu}\hat{\nabla}^{\rangle\nu}\hat{\nabla}^2\hat{\nabla}\cdot\hat{u}+\cdots\nn
 &\vdots~~~~~~
 \label{MIS-type1}
\end{align}
Eq.\eqref{MIS-type1} and so on set a nested series of an infinite number of new degrees of freedom much in the same line as the conventional MIS theory given by,
$T^{\mu\nu}=\varepsilon(u^{\mu}u^{\nu}+\frac{1}{3}\Delta^{\mu\nu})+\pi^{\mu\nu},~(1+\tau_{\pi} D)\pi^{\mu\nu}=-2\eta\sigma^{\mu\nu}$, with $\tau_{\pi}$ as the relaxation time of shear viscous flow. Equations \eqref{MIS-type1} combinedly boil down to Eq.\eqref{BDNK00} and \eqref{BDNK11}, where each increasing spatial gradient term in $\hat{\pi}^{\mu\nu}$ (right-hand side of Eq.\eqref{BDNK11}) is now attributed to a new degree of freedom.
%-----------------------------------------------
%-----------------------------------------------
\section{Frame transformation in one go}
So far we have solved the linearized frame transformation equations \eqref{shifteqn} using order-by-order derivative expansion where the infinite series of field corrections is summed to generate temporal derivatives in the denominator. Now we will use this formal manipulation to have both the spatial and temporal derivatives in the denominator. This will lead to solutions of the frame transformation equations \eqref{shifteqn} in one go contrary to solving them via the chain of recursive relations \eqref{shift-soln}.
The method gives the following two coupled scalar equations in terms of $(\hat\nabla\cdot \delta u)$ and $(\delta T/\hat{T})$ as,
\begin{align}
&\left[1+\tilde{\theta}\hat{D}\right](\hat{\nabla}\cdot\delta u)+\tilde{\theta}\hat{\nabla}^2\frac{\delta{T}}{\hat{T}}+\tilde{\theta}\left[\frac{\hat{\nabla}^2\hat{T}}{\hat{T}}+\hat{D}\hat{\nabla}\cdot\hat{u}\right]=0,\nn
&\left[1+\tilde{\chi}\hat{D}\right]\frac{\delta T}{\hat{T}}+\frac{\tilde{\chi}}{3}(\hat{\nabla}\cdot\delta u)=0~.
\label{onego}
\end{align}
We have used here the on shell identity $\frac{\hat{D}\hat{T}}{\hat{T}}+\frac{1}{3}\hat{\nabla}\cdot\hat{u}=0$ that always holds at the linearized level under Landau frame condition. We first extract $\delta T/\hat T$ by eliminating $(\hat\nabla\cdot\delta u)$ from Eq.\eqref{onego}, and then substituting it in the first equation of \eqref{shifteqn}, we find the expression for $\delta u^\mu$ (details documented in the separate, parallel manuscript). After substituting the velocity correction again in Eq.\eqref{BDNK1}, we get the following expression of shear tensor $\hat{\pi}^{\mu\nu}(=-2\eta[\hat{\sigma}^{\mu\nu}+\partial^{\langle\mu}\delta u^{\nu\rangle}])$ in the Landau frame,
\begin{align}
\hat{\pi}^{\mu\nu}&=
-\left[2\eta\over1+\tilde{\theta}\hat{D}\right]\hat{\sigma}^{\mu\nu}\nn
&+\left[2\eta\tilde{\theta}\over 1+\tilde{\theta}\hat{D}\right]\left[{\left\{1+(\tilde{\theta}+\tilde{\chi})\hat{D}\right\}\frac{\hat{\nabla}^{\langle\mu}\hat{\nabla}^{\nu\rangle}\hat{T}}{\hat{T}}\over (1+\tilde{\theta}\hat{D})(1+\tilde{\chi}\hat{D})-\frac{\tilde{\theta}\tilde{\chi}}{3}~\hat{\nabla}^2}\right].
\label{pi-onego}
\end{align}
The equation \eqref{pi-onego} is a formal solution with spatial as well as temporal derivatives in the denominator. But following the strategy presented before, we can recast equation \eqref{pi-onego} as an inhomogeneous differential equation for the new `nonfluid' degree of freedom $\hat{\pi}^{\mu\nu}$ as follows,
\begin{align}
&\left[(1+\tilde{\theta}\hat{D})(1+\tilde{\chi}\hat{D})-\tilde{\theta}\frac{\tilde{\chi}}{3}~\hat{\nabla}^2\right]
\left\{(1+\tilde{\theta}\hat{D})\hat{\pi}^{\mu\nu}+2\eta\hat{\sigma}^{\mu\nu}\right\}
\nn
&~~~~~~ =~
2\eta\tilde{\theta}\left\{1+(\tilde{\theta}+\tilde{\chi})\hat{D}\right\}\frac{\hat{\nabla}^{\langle\mu}\hat{\nabla}^{\nu\rangle}\hat{T}}{\hat{T}}~.
\label{pi-onegofinal}
\end{align}
Here, just like in MIS theory, we are introducing only one `non-fluid' tensorial degree of freedom, but it follows a complicated inhomogeneous PDE, second order in spatial but third order in temporal derivatives.

So we see, generically a nonlocal theory could be made local by introducing new degrees of freedom, but the process of `integrating in' new degrees could have ambiguities. The two methods described here are one example of this ambiguity. Both methods attempt to write a system of coupled equations involving both fluid and `non-fluid' variables that are equivalent to the equations in BDNK theory. However, the structure of the equations and also the extra `non-fluid' variables are so widely different that in the first case, we need to introduce an infinite number of variables, whereas in the second case, we need just one. However, the field redefinition we have used in the first method (Eq.\eqref{BDNK11}) can be further rearranged in the following fashion,
\begin{align}
 &\hat{\pi}^{\mu\nu}=-2\eta\Bigg[\frac{1}{(1+\tilde{\theta}\hat{D})}\hat{\nabla}^{\mu}\hat{u}^{\nu}+\nn
 &\frac{(-\tilde{\theta})\frac{1}{\hat{T}}\hat{\nabla}^{\langle\mu}\hat{\nabla}^{\nu\rangle}\left\{1+\frac{(-\tilde{\theta})(-\frac{1}{3}\tilde{\chi})\hat{\nabla}^2}{(1+\tilde{\theta}\hat{D})(1+\tilde{\chi}\hat{D})}+\cdots\right\}\hat{T}}{(1+\tilde{\theta}\hat{D})(1+\tilde{\chi}\hat{D})}+\nn
 &\frac{\left(\frac{\tilde{\theta}\tilde{\chi}}{3}\right)\hat{\nabla}^{\langle\mu}\hat{\nabla}^{\nu\rangle}\left\{1+\frac{(-\tilde{\theta})(-\frac{1}{3}\tilde{\chi})\hat{\nabla}^2}{(1+\tilde{\theta}\hat{D})(1+\tilde{\chi}\hat{D})}+\cdots\right\}\hat{\nabla}\cdot\hat{u}}{(1+\tilde{\theta}\hat{D})^2(1+\tilde{\chi}\hat{D})}\Bigg].
 \label{pi-compare}
\end{align}
The infinite sum in powers of the spatial derivative $\hat\nabla^2$ in Eq.\eqref{pi-compare} converges for those linearized perturbations where the operator satisfies the inequality
$\frac{\tilde\theta\tilde{\chi}}{3}\hat{\nabla}^2/[(1+\tilde{\theta}\hat{D})(1+\tilde{\chi}\hat{D})]<1$.
Within this radius of convergence, we can again sum the spatial derivatives and obtain the results of the second method (Eq.\eqref{pi-onego} as well as \eqref{pi-onegofinal}). So we see that, these two sets of equations of defining new 'non-fluid' variables (\eqref{MIS-type1} and \eqref{pi-onegofinal}), are actually equivalent, at least in the regime of frequency and spatial momenta set by the radius of convergence.

At this stage, let us emphasize that the method of `integrating in' new `non-fluid' degrees of freedom with new equations of motion is highly non-unique, even at the linearized level. For example, we could have chosen $\delta u^\mu$ and $\delta T$ themselves to be the new `non-fluid' variables, satisfying the new equations as given in \eqref{shifteqn} and we could take a viewpoint that the $u^\mu$ and the $T$ fields in the BDNK theory are actually the Landau frame fluid variables ($\hat{T},\hat{u}^{\mu}$) plus the `non-fluid' variables $\{\delta u^\mu, \delta T\}$. Note that though $\delta u^\mu$ and $\delta T$ would look very much like the velocity and temperature corrections, they are still `non-fluid' variables in the Landau frame since they vanish in global equilibrium. Another choice of introducing infinitely many `non-fluid' degrees of freedom would be to simply use $\delta u^\mu_n$ and $\delta T_n$ (as defined in \eqref{shift-order})  and then the recursive equations \eqref{shift-soln} would turn out to be the new equations of motion.

%=================================================
%------------------------------------------------
%------------------------------------------------
\section{Conclusion}
In this letter, we rewrite the stress tensor of the BDNK hydrodynamic theory in the Landau frame at least for the part that will contribute to the spectrum of linearized perturbation around static equilibrium. Though the BDNK formalism has a finite number of derivatives, it turns out that in the Landau frame, it will have either an infinite number of derivatives or one has to introduce new non-fluid variables. There is no unique way to introduce these non-fluid variables. Here, we have presented two different ways of doing it, resulting in two completely different-looking sets of equations. The two choices of new variables, discussed here, are basically guided by our sense of mathematical aesthetics and an attempt to adhere to the philosophy of the MIS theory where the new `non-fluid' variable is a rank-2 symmetric tensor, structurally very similar to the energy-momentum tensor.

Our work has set up a stage for comparison between the BDNK and the MIS-type theories or their respective predictions about the fluid evolution. At first glance, they look very different. But the fluid variables like velocity and temperature used to express the BDNK stress tensor are not the same as the ones used in the MIS theory. A comparison is meaningful only if the basic variables of the equations are the same. Once we have done the required transformation (general frame to Landau frame), it turns out that though there are differences in the details, the basic structure of nonlocality or `non-fluid' variables is very similar in both the theories. The advantage of the Landau frame is that the fluid variables are locally defined in terms of the one-point function of the stress tensor, and in this case, the obtained causal system of equations (after frame transformation), turns out to have nonlocal terms that are 'integrated in' as the new degrees of freedom. Whereas in the original BDNK theories, the equations are local with a finite number of derivatives, but the fluid variables are related to the one-point function of the stress tensor in a very non-trivial and nonlocal fashion.
However, there is more information in the BDNK formalism than what has just been stated above. It says that there exist causal fluid theories where the non-localities could be completely absorbed in a field redefinition, thereby generating causal but local fluid theory with a finite number of derivatives. Since the final equation we derived for the shear tensor $\hat{\pi}^{\mu\nu}$ is not exactly the same as in the conventional MIS theory, it also says that the non-localities of the MIS theory can possibly never be completely absorbed in the field redefinition.

It would be interesting to extend this analysis to full nonlinear order. Also, it would be very informative to know whether and, if yes, how the story changes as one adds higher-order derivative corrections to the BDNK theory.

%=================================================
\section{Acknowledgements}

S.B. acknowledges B. Withers for helpful discussions and Trinity College, Cambridge for hospitality while this manuscript was being prepared.
S.B., S.M. and S.R. acknowledge the Department of Atomic Energy, India, for the funding support.

%==================================================

\end{document}